\documentclass[12pt]{JHEP3}
\usepackage{amssymb}
\usepackage{bbold}
\usepackage{graphicx}
\newcommand{\1}{\mathbb 1}
\title{\bf P-term Strings and Semi-local Strings}
\author{C. Burrage\\ E-mail:  \email{C.Burrage@damtp.cam.ac.uk}\\Department of Applied Mathematics and Theoretical Physics\\
\ Centre for Mathematical Sciences\\
 Cambridge CB2 0WA, United Kingdom} 
\author{A.C. Davis\\ E-mail:  \email{A.C.Davis@damtp.cam.ac.uk}\\ Department of Applied Mathematics and Theoretical Physics\\
 Centre for Mathematical Sciences\\
 Cambridge CB2 0WA, United Kingdom} 
  \abstract{
P-term potentials can give rise to Nielsen-Olesen or semi-local
cosmic strings.  We present a general analysis of these cosmic strings
where we derive the Bogomol'nyi equations and field profiles for
both types of string and discuss their stability.  We give an analysis
of the
fermionic zero modes that could live on the strings and a brief
discussion of the inflationary period preceding their formation.}

\begin{document}
\section{Introduction}

There has been much interest in obtaining cosmological inflation 
from string theory. Generically such attempts have
problems obtaining a Minkowski vacuum and stabilising the moduli
fields. The KKLMMT scenario \cite{Kachru:2003aw,Kachru:2003sx} stabilises 
the volume modulus and it is therefore interesting to study the properties of 
this model. Typically brane inflation models give rise to lower dimensional
branes, which are formed at the end of inflation, with the formation of cosmic
strings being generic \cite{Majumdar:2002hy,Sarangi:2002yt}.
In this paper we will study the cosmic strings that form at the end of 
inflation, for a review see \cite{Polchinski:2004ia,Davis:2005dd}.

The D3/D7 version of the KKLMMT model \cite{Dasgupta:2004dw,Dasgupta:2002ew} gives rise to a potential for
the bosonic scalars known as a P-term potential.  This is an N=2
supersymmetric potential constructed from the triplet of auxiliary
fields $P_i$ which are given a vev by a triplet of Fayet-Iliopoulos
(FI) terms $\xi_i$.  This potential has the property that if we
truncate to N=1 supersymmetry the potential can look like a D-term
potential, an F-term potential or a mix of the two depending on the
direction of $\xi_i$.  P-term models were introduced in \cite{Kallosh:2001tm} with a
global SU($2,2|2$) superconformal gauge theory where they correspond to a
dual gauge theory of supersymmetric D3/D7 branes \cite{Kallosh:2003ux}.  This is broken to
N=2 supersymmetry by giving a vev to $P_i$, which corresponds to a
magnetic flux triplet in the D3/D7 brane construction.

It is not straightforward to construct a P-term model in
supergravity.  It was thought that it was not possible to include the
triplet of FI terms in N=2 supergravity, however in \cite{Burrage:2007yu} we
demonstrated that this could be done.  In this paper we construct
the potential in the more familiar language of N=1 supersymmetry
where the second supersymmetry arises for a particular tuning of the
parameters.  However local N=1 supersymmetry with an FI term requires
an R-symmetry; the superpotential must be invariant under the
R-symmetry and charged under the U(1) associated with the FI term \cite{Freedman:1977,Stelle:1978wj,Barbieri:1982ac,Binetruy:2004hh}.
This means the charges of the scalar fields in the superpotential must
be different in supergravity to their supersymmetric values which has
consequences for the construction of P-term models in supersymmetry, as
we will see in section \ref{model}.

This paper examines the topological defects that form in P-term
potentials,  in particular whether they are permitted by current
cosmological constraints.  Simple P-term potentials have a supersymmetric
minimum with an $S^1$ degeneracy which means that Nielsen-Olesen (NO)
cosmic strings can form by the Kibble mechanism \cite{Kibble:1976sj}, however 
such strings typically have too high a string tension to agree with 
observations.  We also consider the P-term model when a second set of charged
chiral multiplets are included such that there is
an SU(2) global symmetry between the scalar multiplets.  This is known
as the semi-local model.  Semi-local 
strings are not topologically stable and as a result they do not
conflict with observations.  Semi-local strings appear when global and gauge
symmetries coexist in a model,  they were shown to arise in the
developed D3/D7 brane inflation model \cite{Dasgupta:2004dw}, where the 
effective potential was P-term.

Section \ref{model} reviews the P-term model in supersymmetry and how
the charges of the fields change when we move to supergravity.  In
section \ref{strings} we find the general P-term form of the  Bogomol'nyi equations which
determine the field profiles for NO and semi-local strings.  We find the asymptotic behaviour
of both types of string at the core of the string and also at
infinity.  From this we find the string tension which can be compared
with the allowed value determined by observations.  In section
\ref{SL} we also discuss the stability properties of semi-local
strings which mean that they can avoid this bound.

In section \ref{zero} we discuss the zero modes that can exist on the
strings.  We give the supersymmetry transformations for all the
fermions and then in \ref{indextheorem} we discuss the index theorem
for counting fermionic zero modes in supergravity \cite{Brax:2006yb}.  In \ref{zeroD} we discuss D-term strings which are an
exceptional case because of their BPS nature, and in \ref{zeroF} we
discuss F-term strings and generic P-term strings. 
In section \ref{inflation} we discuss inflation in these potentials.
Again because of their different properties we treat the generic
P-term case and the D-term case in
sections \ref{Pinflation} and  \ref{Dinflation} respectively. We conclude in section \ref{conc}.

\section{The P-term model}
\label{model}
We begin with a description of the P-term potential in the language of
N=1 supersymmetry.
Take a theory which contains three chiral
multiplets with charges 0,+1,-1,  and a
vector multiplet and arrange the two charged scalar fields into a
multiplet of charge $+1$
\begin{equation}
\label{multiplet}
h=\left(\begin{array}{c}
\phi_{+}\\
\phi_{-}^*
\end{array}\right)
\end{equation}
and define
\begin{equation}
\label{P}
P_i = h^{\dagger}\sigma_ih-\xi_i
\end{equation}
$P_i$ are the triplet of auxiliary fields, or moment maps, of the
  N=2 theory and $\xi_i$ is a constant FI three vector \cite{Kallosh:2001tm}.  The
  superpotential and D-term are 
\begin{eqnarray}
W&=&\frac{\beta\phi_0}{2}(P_1-iP_2)\nonumber\\
&=& \beta\phi_0\left(\phi_+\phi_--\frac{1}{2}(\xi_1-i\xi_2)\right)\label{sup}\\
D&=&\frac{g}{2}P_3\nonumber\\
&=&\frac{g}{\sqrt{2}}(|\phi_+|^2-|\phi_-|^2-\xi_3)\label{D}
\end{eqnarray}
where $\beta=g/\sqrt{2}$.  Tuning the parameters in this way makes the
masses of the vector and scalar particles equal, and it is in this
limit that the second supersymmetry, which is anti-chiral, emerges \cite{Achucarro:2004ry}.
A P-term potential in N=1 supersymmetry contains constant FI terms in the
 superpotential and the D-term.

 The scalar potential is  constructed from the superpotential and D-term as $V=|\partial W|^2+\frac{g^2}{2}D^2$
\begin{equation}
V=\frac{g^2}{8}\sum_{i=1}^3(h^{\dagger}\sigma_i h - \xi_i)^2 +
\frac{g^2}{2}|\phi_0|^2 h^{\dagger}h
\label{Spotential} 
\end{equation}
 $\phi_0$ is the scalar in the uncharged chiral multiplet, $g$ is a
gauge coupling constant, and $\sigma_i$ are the Pauli matrices.  If
$\xi_1 = \xi_2 = 0$
this potential is the super-Bogomol'nyi limit\footnote{as
  defined in \cite{Achucarro:2004ry}} of a D-term potential; if
$\xi_2 = \xi_3 = 0$ this is the Bogomol'nyi limit of an
F-term potential.  The field profiles and zero modes of F- and D-term
cosmic strings in supersymmetry are discussed in \cite{Davis:1997bs}.  We split the three vector $\xi_i$ into the
product of a rotational part described by a SO(3) matrix $R$
and a magnitudinal part $(0,0,\xi)$
\begin{equation}
\label{xii}
\xi_i = (R^{-1})_{ij}\delta_{j3}\xi
\end{equation}
 so all P-term
potentials are described by rotations from the D-term case.
 The class of P-term potentials is parameterised by two Euler
 angles; $\psi$, $\theta$, and the magnitude of the FI vector; $\xi$.
 Our conventions for Euler angles are given in appendix \ref{angles}.

The semi-local model includes a second pair of charged
scalars such that there is an SU(2) global symmetry amongst
the charged fields.  The supersymmetric scalar potential becomes
\begin{equation}
\label{SLpot}
V=\frac{g^2}{8}\sum_{i=1}^3(h^{\dagger}\sigma_i h+\tilde{h}^{\dagger}\sigma_i \tilde{h} - \xi_i)^2 +
\frac{g^2}{2}|\phi_0|^2( h^{\dagger}h+\tilde{h}^{\dagger}\tilde{h})
\end{equation}
Semi-local strings have different stability properties to NO cosmic
strings, and they therefore avoid some of the cosmological
constraints that NO strings are subject to. 

The potentials (\ref{Spotential}) and (\ref{SLpot}) have two types of minima; one when $|\phi_0|=0$ and the charged fields take fixed
vevs, the other when the charged fields vanish and
$|\phi_0|^2>\xi/2$.  The first minimum preserves the supersymmetry the
second breaks it.  This is the vacuum structure required for hybrid
inflation \cite{Linde}.

After lifting the theory to supergravity we want to be able to make
comparisons with observations, hence we embed the P-term model in N=1
supergravity ignoring the second supersymmetry \cite{Kallosh:2003ux}.  If a  U(1) gauge theory with an FI
term is to be consistently coupled to N=1 supergravity in a way that preserves gauge
invariance the superpotential must be invariant under the R-symmetry
but transform under the U(1) symmetry.  This alters the charges of the fields appearing in the superpotential.  In
supersymmetry the scalar fields $\phi_{\pm}$, $\phi_0$ have
charges $Q_{\pm}=\pm1$, $Q_0=0$, and the superpotential
is uncharged.  In the supergravity version of the P-term model the fields have charges
\begin{eqnarray}
\label{charges}
q_i=Q_i-\rho_i\frac{\xi}{M_P^2} &, & \sum \rho_i=1
\end{eqnarray}
As $M_P\rightarrow \infty$ we regain $q_i=Q_i$, indeed with a generic choice of superpotential the charges always return
to their supersymmetric values in this limit.  $M_P \rightarrow \infty$ is known as the rigid limit of supergravity
\cite{Binetruy:2004hh} and describes supersymmetry in a curved space-time.  Unless the rigid limit is being considered it no longer makes sense to combine $\phi_+$,
$\phi^*_-$ into the multiplet (\ref{multiplet}).
A P-term model in N=1 supergravity has the following 
superpotential and D-term
\begin{equation}
\label{Wsugra}
W=\frac{g\phi_0}{\sqrt{2}}\left(\phi_+\phi_--\frac{1}{2}(\xi_1-i\xi_2)\right)
\end{equation}
\begin{equation}
\label{Dsugra}
D=\frac{g}{\sqrt{2}}(q_0|\phi_0|^2+q_+|\phi_+|^2+q_-|\phi_-|^2-\xi_3)
\end{equation}
P-term potentials were
constructed in \cite{Burrage:2007yu} directly in N=2 supergravity
where each component of the moment map $P_i$ contained a constant term.
Following \cite{Achucarro:2005vz} this was done by considering the truncation to N=1 supergravity where
FI terms were present in the D-term and
superpotential of the reduced N=1 supergravity theory.  Terms involving
the uncharged scalar field were found in the D-term as a result of the non-Abelian gauging of the N=2 theory needed to
produce a P-term potential.  This should be compared with the presence
of the $q_0|\phi_0|^2$ term in (\ref{Dsugra}).

\section{Field Profiles for P-term Strings}
\label{strings}
Cosmic strings are one dimensional topological defects which 
form at the end of hybrid inflation for P-term potentials.  We shall consider the form of NO
strings, which form when there are two charged chiral multiplets in the
model, and semi-local strings, which form when a second set of charged
chiral fields are included.  In
our discussion of cosmic strings we will use the supersymmetric P-term
potential as the vevs of the scalar fields are
always very far from the Planck scale so the supergravity corrections to
the potential are negligible.

\subsection{NO Strings}
\label{NO}
We present here a unified formalism for NO strings forming in P-term potentials.
 The
most general metric for a straight, static string  \cite{VilShel}  is
\begin{equation}
\label{metric}
ds^2 = dt^2-dz^2 -dr^2-C^2(r)d\theta^2
\end{equation}
The string energy integral can be written as a sum of positive semidefinite terms and a surface integral term \cite{Comtet:1987wi},
\begin{eqnarray}
\mu_{\tiny \mbox{string}}=
 \int\,drd\theta \;C & &\left(\frac{1}{2}\left|\left(D_r+
i\frac{1}{C}D_\theta\right)(Uh)_1\right|^2\right.\nonumber\\
& &+\frac{1}{2}\left|\left(D_r+i\frac{1}{C}D_\theta\right)(Uh)_2^*\right|^2\nonumber\\
& &+\frac{1}{2} (B-\frac{g}{2}(h^{\dagger}U^{\dagger}\sigma^3Uh-\xi))^2\nonumber\\
& &+ \frac{g^2}{8}(h^{\dagger}U^{\dagger}\sigma^2Uh)^2 +
\frac{g^2}{8}(h^{\dagger}U^{\dagger}\sigma^1Uh)^2\nonumber\\
& &+\left.\frac{g^2}{2}|\phi_0|^2|Uh|^2 +
\frac{1}{C}(\underline{\nabla}\times\underline{J})_z\right)
\label{stringenergy}
\end{eqnarray}
where the last term can be rewritten as a surface integral at
infinity. $D_{\mu}$ is a covariant derivative $D_{\mu}\phi_i = \left(\partial_{\mu}-igQ_iA_{\mu}\right)\phi_i$, $A_{\mu}$ is the gauge potential, $F_{\mu\nu}$ is the
corresponding gauge field, and $B=F_{12}$.
\begin{equation}
J_{\mu} =\frac{1}{2}i\Bigl((D_{\mu}(Uh)^{\dagger})(Uh) -
(Uh)^{\dagger}D_{\mu}(Uh)\Bigr) + gA_{\mu}\xi
\end{equation}
and  $U$ is the SU(2) rotation related to the SO(3) rotation $R$, which is
given in terms of Euler angles in appendix \ref{angles}.  
It is possible to relate $J_{\mu}$ to the gravitino potential; by choosing the trivial K\"{a}hler potential
\begin{equation}
\label{kahler}
K= \frac{|\phi_+|^2+|\phi_-|^2 + |\phi_0|^2}{M_P^2}
\end{equation}
the gravitino potential is $A_{\mu}^B = J_{\mu}/M^2_P$.
 The minimum
value for $\mu_{\tiny \mbox{string}}$ is
obtained when each of the squared terms in (\ref{stringenergy})
vanishes, these conditions are know as the Bogomol'nyi equations.  In
 particular they require
\begin{equation}
\label{secondiszero}
(Uh)_2 \equiv 0
\end{equation}
If the Bogomol'nyi equations are satisfied the Einstein equation becomes
\begin{equation}
C^{\prime}+\frac{1}{M_p^2}J_{\theta} = \mbox{const}
\end{equation}

The most general form of $h$ for a straight, static, infinite cosmic
string lying along the z axis is
\begin{equation}
\begin{array}{ccc}
h&\equiv \left( \begin{array}{c}
\phi_+\\
\phi_-^*
\end{array} \right)
&= e^{in\theta}\left( \begin{array}{c}
f(r)\\
e^{i\Delta}h(r)
\end{array} \right)
\end{array}
\end{equation}
where $f(r)$ and $h(r)$ are real functions of the radial coordinate and
$\Delta$ is a real constant.  Without loss of generality we take $n$, the
winding number of the string, to be
positive. Thus  (\ref{secondiszero}) gives
\begin{equation}
ah(r)=be^{i\Delta}f(r)
\end{equation}
where $a=e^{i(\phi+\psi)/2}\cos(\theta/2)$ and
$b=ie^{i(\phi-\psi)/2}\sin(\theta/2)$ are Cayley-Klein parameters of
the rotation. 
The most general form of the gauge potential is
\begin{eqnarray}
\label{gauge}
A_r=0, & & A_{\theta}=\frac{n\alpha(r)}{g}
\end{eqnarray}
so that $B=n\alpha^{\prime}(r)/gC(r)$.  The boundary conditions for these fields are  
\begin{equation}\begin{array}{cc}
C(0)=0 & C^{\prime}(0)=1\\
f(0)=0 & f(\infty)=|a| \xi^{\frac{1}{2}}\\
\alpha(0)=0 & \alpha(\infty)=1
\end{array}
\end{equation}
to ensure that the string energy is finite at infinity and non
singular at the origin.  In terms of these fields the Bogomol'nyi equations describing the minimum string energy field
configuration are
\begin{equation}
f^{\prime} -\frac{n}{C}(1-\alpha)f = 0
\label{mol1}
\end{equation}
\begin{equation}
\frac{n\alpha^{\prime}}{gC} = \frac{g}{2}\left( \frac{f^2}{|a|^2}
-\xi\right)
\label{mol2}
\end{equation}
and the Einstein equation
becomes
\begin{equation}
1=C^{\prime} +
\frac{nf^2(1-\alpha)}{M_P^2|a|^2}+\frac{n\alpha \xi}{M_P^2}
\end{equation}
It is straightforward to rewrite these equations if $a=0$.  There are no known exact solution to these equations
in the general case but the behaviours of the fields in
the small and large $r$ limits can be examined:
As $r \rightarrow \infty$
\begin{eqnarray}
f(r) &\rightarrow& |a|\sqrt{\xi}\\
\alpha(r) &\rightarrow& 1\\
C(r) &\rightarrow& \left( 1-\frac{n\xi}{M_P^2} \right)r
\end{eqnarray}
and close to the origin
\begin{eqnarray}
C(r) &=& r+\mathcal{O}(r^2)\\
f(r)&=& \beta_nr^n +\mathcal{O}(r^{n+1})\\
\alpha(r)& =&  \frac{-\xi g^2}{4n}r^2 + \mathcal{O}(r^3)
\end{eqnarray}
where $\beta_n$ is a constant.

If the fields are in the Bogomol'nyi configuration we can compute
the minimum energy of the string.  The resulting string tension is
\begin{equation}
\label{stringtension}
G\mu_{\tiny \mbox{string}}=\frac{n\xi}{4M_P^2}
\end{equation}
for all P-term models.  Current estimates of the string tension give
$G \mu \sim 10^{-7}$ \cite{Copeland:2003bj}, which would mean that
\begin{eqnarray}
\xi &\sim & 10^{-7}M_P^2
\end{eqnarray}
if $n$ is of order one. We will see in section \ref{inflation} that
this disagrees with the bounds on $\xi$ coming from observations of
the angular power spectrum.  

\subsection{Semi-Local Strings}

\label{SL}
The potential for semi-local strings was given in equation
(\ref{SLpot}), in what follows we give a unified description of
semi-local strings in P-term potentials.  Repeating the same analysis as for the NO case we find that the field profiles are very similar to those of
the NO string.  However semi-local strings are unstable due to a degeneracy
in the equations, and so unlike NO strings we would not expect any
semi-local strings formed at the end of inflation to have survived
long enough to affect cosmological observations.

The string energy in the semi-local model takes the same form as in
the NO case, but with the semi-local potential (\ref{SLpot}), and with kinetic
terms for the tilded charged scalar fields.  Performing the same rearrangement as in the NO case produces a
set of Bogomol'nyi equations for the fields:
\begin{eqnarray}
\left(D_r +i\frac{1}{C}D_{\theta}\right)(Uh)_1 =0 & & \left(D_r
+i\frac{1}{C}D_{\theta}\right)(U\tilde{h})_1 =0\\
\left(D_r +i\frac{1}{C}D_{\theta}\right)(Uh)_2^{*} =0 & & \left(D_r
+i\frac{1}{C}D_{\theta}\right)(U\tilde{h})_2^{*} =0
\end{eqnarray}
\begin{equation}
B-\frac{g}{2}(h^{\dagger}U^{\dagger}\sigma^3U h+
\tilde{h}^{\dagger}U^{\dagger}\sigma^3U\tilde{h} - \xi) =0
\end{equation}
\begin{equation}
\begin{array}{ccc}
(Uh)_2=0, & (U\tilde{h})_2=0, & |\phi_0|=0 \label{h2zero}
\end{array}
\end{equation}
The Einstein equation is
\begin{equation}
C^{\prime}+A_{\theta}^B = 1
\end{equation}
where the constant is determined by the
boundary conditions.
The most general form of
the bosonic profiles for a straight cosmic string satisfying (\ref{h2zero}) is
\begin{equation}
\begin{array}{lr}
h=\frac{e^{in\theta}f_1(r)}{a^*}\left(\begin{array}{c}
a^*\\
b^{*}
\end{array}\right)
& \tilde{h}=\frac{e^{im\theta}f_2(r)}{a^*}\left(\begin{array}{c}
a^*\\
b^{*}
\end{array}\right)
\end{array}
\end{equation}
$m$ is an arbitrary integer and  without loss of generality
  we set $m \leq n$ so that $n$ is the winding number of the string.  The form of the gauge field is given in (\ref{gauge}). Again
  it is straightforward to rewrite the equations
  if $a=0$.
  The Bogomol'nyi and Einstein equations  become;
\begin{eqnarray}
\left( \partial_r -\frac{n}{C}(1-\alpha)\right)f_1 &=& 0 \label{SLbog1}\\
\left(\partial_r -\frac{1}{C}(m-n\alpha)\right)f_2 &=& 0 \label{SLbog2}\\
\frac{n\alpha^{\prime}}{gC}- \frac{g}{2}\left(
\frac{f_1^{\;2}}{|a|^2}+\frac{f_2^{\;2}}{|a|^2} -\xi \right) &=& 0\\
\frac{1}{M_P^2|a|^2}(f_1^{\;2}(n-n\alpha)+f_2^{\;2}(m-n\alpha))+\frac{n\alpha
  \xi}{M_P^2} &=& 1-C^{\prime}
\end{eqnarray}
To ensure the finiteness and regularity of the string energy the boundary conditions are
\begin{eqnarray}
C(0)=0 & & C^{\prime}(0)=1\\
f_1(0) =0 & & f_2(0)=f_0\delta_{m0}\\
\alpha(0)=0 & & \alpha^{\prime}(\infty)=0\\
& &f_1^{\;2}(\infty)+f_2^{\;2}(\infty)=|a|^2\xi
\end{eqnarray}
where $f_2$ is allowed to be non-zero at the origin if $m=0$, which is a scalar condensate at the core of the
string.
However $f_1(r)$ and $f_2(r)$ are not independent functions,
(\ref{SLbog1}) and (\ref{SLbog2}) can be combined to give
\begin{equation}
\label{mnbounds}
\frac{\partial}{\partial r}\ln \left( \frac{f_2}{f_1}\right) =
\frac{m-n}{C}
\end{equation}

For the Bogomol'nyi equations to hold with $m\leq n$
equation (\ref{mnbounds}) requires that as $r \rightarrow \infty$
\begin{eqnarray}
f_1^2(r) &\rightarrow& \xi|a|^2\\
f_2^2(r)&\rightarrow& 0
\end{eqnarray}
and hence
\begin{eqnarray}
\alpha(r)&\rightarrow&1\\
C(r) &\rightarrow& r\left(1-\frac{n\xi}{M_P^2}\right)
\end{eqnarray}
Notice that the form of the metric at large $r$, and hence the
deficit angle of the string, is unchanged from that of an NO
string.
For the fields to be finite at the origin requires $m,n\geq0$.  At
small $r$ 
\begin{eqnarray}
f_1(r)&=& \beta_nr^n + \mathcal{O}(r^{n+1})\\
f_2(r) &=& \gamma_mr^m + \mathcal{O}(r^{m+1})\\
C(r) &=& r + \mathcal{O}(r^2)\\
\alpha(r) &=& \frac{-\xi g^2}{4n}r^2+\mathcal{O}(r^3)
\end{eqnarray}

If the fields are in the Bogomol'nyi configuration the string tension is
\begin{equation}
G\mu_{\tiny \mbox{string}}= \frac{ \xi n}{4M_P^2}
\end{equation}
which is the same as the NO case (\ref{stringtension}).  However
this does not conflict with observations as semi-local strings are unstable.  As shown in  \cite{Hindmarsh:1992yy,Penin:1996si,Achucarro:1999it} for any mode with $m<n$ there is a degeneracy
in the solutions to the Bogomol'nyi equations; that is the solutions to
(\ref{mnbounds}) are a one parameter
family of defects which  all have the same energy.  Any solution can
interpolate between an NO string and a $\mathbb{C}P^1$ lump with no
cost in energy.  In other words the flux along the string is not
confined within a tube of any particular radius and for any semi-local
string generic perturbations will excite degenerate modes which will
force the size of the flux tubes to ever larger values.  Therefore
although semi-local strings may have formed at the end of inflation,
they would have been transitional and would have rapidly decayed
away. 

This degeneracy was invoked in \cite{Urrestilla:2004eh} to produce a
model of D-term inflation which did not give rise to cosmic strings.
The same property means that a model of P-term inflation can be constructed
which also does not conflict with observations.  Infinite semi-local
strings were expected to be unlikely to form in our universe for a
different reason in \cite{Dasgupta:2004dw}.  Here it was thought that
the probability of producing an infinite semi-local string from short
segments would be small.  

\section{Zero Modes}
\label{zero}
Fermionic zero modes are a common feature of cosmic strings in
supersymmetry and they alter the resulting cosmology of a model because they give
rise to currents on the string.  The fermions present in our system are four charged chiral fields $\chi_+$, $\chi_-$,
$\tilde{\chi}_+$ and $\tilde{\chi}_-$, a chargeless chiral field $\chi_0$, a
gaugino $\lambda$ and a gravitino $\psi_{\mu L}$.  The supergravity
transformations of the fermions are
\begin{eqnarray}
\delta(\chi_{\pm}) &=& \frac{1}{2}\left(\sigma^rD_r
+\frac{1}{C}\sigma^{\theta}D_{\theta}\right)\phi_{\pm}\bar{\epsilon}\\
\delta(\tilde{\chi}_{\pm}) &=& \frac{1}{2}\left(\sigma^rD_r
+\frac{1}{C}\sigma^{\theta}D_{\theta}\right)\tilde{\phi}_{\pm}\bar{\epsilon}\\
\delta\chi_0 &=& \frac{-g}{2}((P_1)^2 +
(P_2)^2)^{\frac{1}{2}} \1 \epsilon\\
\delta\lambda &=& i\left(\sigma_3B + \frac{g}{2}P_3 \1 \right)\epsilon\\
\delta \psi_{\mu L} &=& \left(\partial_{\mu} +
\frac{1}{4}\omega_{\mu}^{ab}\sigma_{ab} + \frac{1}{2}iA_{\mu}^B
\right) \epsilon_L
\end{eqnarray}
where $P_{\mu}$ is defined in equation (\ref{P}).  
 Note that if $U=\1$, which gives D-term strings, the system is
 half-BPS; half of the fermion transformations vanish.

In supersymmetry, when the gravitino is absent, it has been shown
\cite{Brax:2006yb} that generic D-term strings have $2n$ modes
of positive chirality and no modes of negative chirality.    However D-term strings are
normally studied away from the Bogomol'nyi limit.
In the P-term model there are actually two supersymmetries
present \cite{Achucarro:2004ry}, the second supersymmetry is anti-chiral and gives $2n$ zero
modes of negative chirality and no modes of positive chirality.  So in
total in the Bogomol'nyi limit where there are two supersymmetries
present a D-term model has $2n$ zero modes of each chirality.  This is
expected as N=2 supersymmetry is not chiral. F-term
strings have $2n$ modes of each chirality, and so we would expect to find $2n$ zero modes of each chirality
for a P-term string in supersymmetry.

\subsection{The index theorem}
\label{indextheorem}
We will use the index theorem of \cite{Brax:2006yb} to calculate the
number of fermionic zero
modes of P-term cosmic strings in N=1 supergravity.  An important
consideration is whether or not the strings are BPS, as this changes
the way the index theorem is calculated because the mass
matrix becomes block off diagonal.  Therefore when counting the zero
modes of P-term cosmic strings we have to consider the D-term case
separately.  In what follows we describe how to compute the number of
zero modes for BPS strings as an example of how the index theorem is formulated.  The general case proceeds in a
similar way and for full details we refer the reader to
\cite{Brax:2006yb}.   

The number of zero modes of positive and negative chirality is
calculated by considering the normalisability of the fermionic zero
modes at the origin and at infinity.  In the BPS case the Dirac mass
matrix is block off diagonal with entries in the $n_1 \times n_2$
upper right corner and $n_2 \times n_1$ lower left corner of the matrix.  The fermions diagonalise the string
generator such that $T_s\chi^a=q_a\chi^a$.  If $q_a$ is the charge of the a-th
fermion then we write $q_a^{(1)}=q_a$ for $a=1 \dots n_1$  and
$q_a^{(2)}=q_{a+n_1}$ for $a=1 \dots n_2$.  We chose  $\eta$ so
that $q_a^{(1)}+\eta \in \mathbb{Z}+1/2$ and
$q_a^{(2)}-\eta \in \mathbb{Z}+1/2$.  The number
of massless fermions at infinity is $n_z$, the number of
massless fermions with $a \leq n_1$ is $n_{z1}$ and the number of
massless fermions with $a>n_1$ is $n_{z2}$. The number of massive fermions is
$2\bar{n}$ where $\bar{n}=n_1-n_{z1}=n_2-n_{z2}$. 

Given a massless fermion with index $a$ which is not the gravitino we define
\begin{equation}
\tilde{q}^{(1)}_{\pm}=\pm\frac{1}{2}-\eta\mp\left[\frac{C_1}{2}\mp
  \eta\right]
\end{equation}
if $a>n_1$, or
\begin{equation}
\tilde{q}^{(2)}_{\pm} = \pm\frac{1}{2}+\eta \mp \left[\frac{C_1}{2}
  \pm \eta\right]
\end{equation}
if $a\leq n_1$, where $[x]$ is the lowest integer which is strictly greater than $x$.  $C_1 = 1-n\xi$ and
is related to the deficit angle of the string which is given by $\delta=2\pi(1-C_1)$.

After gauge fixing the
gravitino field has three components, for a cosmic string background
it is convenient to write them in terms of three independent Weyl
fermions
\begin{equation}
\begin{array}{ccc}
\Sigma = \sigma^r\bar{\psi}_r+\sigma^{\theta}\bar{\psi}_{\theta}, &
\Psi= \sigma^r\bar{\psi}_r-\sigma^{\theta}\bar{\psi}_{\theta}, & 
\Pi = \sigma^t\bar{\psi}_t-\sigma^z\bar{\psi}_z
\end{array}
\end{equation}
To write the equations of motion for the gravitino in a form suitable
for this analysis take
\begin{equation}
\begin{array}{ccc}
q^{(2)}_{\Psi}=q_{\psi}\mp1, & q^{(2)}_{\Sigma}= q_{\psi}\pm1, & q^{(1)}_{\Pi}=-q_{\psi}\pm1
\end{array}
\end{equation}
where $q_{\psi}=-n\xi/2$.
For a BPS configuration the gravitino is massless so we define 
\begin{eqnarray}
\tilde{q}^{(1)}_{\Psi \pm}&=& \pm \frac{1}{2}-\eta \mp \left[
  \frac{-C_1}{2} \mp \eta \right]\\
\tilde{q}^{(1)}_{\Sigma \pm}&=& \pm \frac{1}{2} -\eta \mp \left[
  \frac{3C_1}{2} \mp \eta \right]
\end{eqnarray}
By considering the field equations for the gravitino it can be seen that
$\Pi$ is pure gauge and decouples from the other fields so we can
ignore it.

We define $\hat{q}_a^{(1)}$ to be the set of $q_a^{(1)}$
and $n_{z2}$ copies of $\tilde{q}^{(1)}_{\pm}$ including
$\tilde{q}^{(1)}_{\Psi \pm}$ and $\tilde{q}^{(1)}_{\Sigma \pm}$, 
ordered so that $\hat{q}^{(1)}_1 \leq \dots \leq
\hat{q}^{(1)}_{\bar{n}+n_z}$.  Similarly $\hat{q}_a^{(2)}$ are
the set of the $q^{(2)}_a$ and $n_{z1}$ copies of
$\tilde{q}^{(2)}_{\pm}$ ordered so that $\hat{q}^{(2)}_1 \geq \dots
\geq \hat{q}^{(2)}_{\bar{n}+n_z}$.  The number of zero modes of
positive and negative chirality is then given by
\begin{equation}
N^{\pm}=2\sum_{a=1}^{\bar{n}+n_z} [\pm
  \hat{q}^{(1)}_a\pm\hat{q}^{(2)}_a]_+
\end{equation}
where $[x]_+=x$ if $x\geq0$ and zero otherwise.

\subsection{D-term strings}
\label{zeroD}
D-term strings are BPS states so we can apply the index theorem just
described.  For an NO string the index theorem says that there are $2(n-1)$ zero
modes of positive chirality and no zero modes of negative chirality in
supergravity \cite{Brax:2006yb}.
 It was
suggested that the vanishing of two of the zero modes compared to the
SUSY result could be a super Higgs effect \cite{Jeannerot:2004bt}. 

To find the corresponding results for semi-local strings we consider
positive and negative chirality modes separately. 
The field $\Sigma$ decouples for positive chirality modes, so we
need only consider the fields $\chi_+$, $\chi_-$, $\tilde{\chi}_+$,
$\tilde{\chi}_-$, $\chi_0$, $\lambda$, $\Phi$.  We have $n_1=4$, $n_2=3$, $n_{z1}=3$ and $n_{z2}=2$.
Then
\begin{equation}
\begin{array}{lllllll}
\hat{q}_1^{(1)}&=\tilde{q}_{+}^{(1)}&=-1-q_{\psi},& &\hat{q}_1^{(2)}&=q_{\chi_0}&=-q_{\psi}\\
\hat{q}_2^{(1)}&=q_{\chi_-}&=-q_{\psi},& &\hat{q}_2^{(2)}&=q_{\lambda}&=q_{\psi}\\
\hat{q}_3^{(1)}&=q_{\tilde{\chi}_-}&=-q_{\psi}, & &\hat{q}_3^{(2)}&=\tilde{q}_+^{(2)}&=q_{\psi}\\
\hat{q}_4^{(1)}&=\tilde{q}^{(1)}_{\Psi +}&=-q_{\psi},& &\hat{q}_4^{(2)}&=\tilde{q}^{(2)}_+&=q_{\psi}\\
\hat{q}_5^{(1)}&=q_{\tilde{\chi}_+}&=m-q_{\psi},& &\hat{q}_5^{(2)}&=\tilde{q}_+^{(2)}&=q_{\psi}\\
\hat{q}_6^{(1)}&=q_{\chi_+}&=n-q_{\psi},& &\hat{q}_6^{(2)}&=q_{\Psi}&=-1+q_{\psi}\\
\end{array}
\end{equation}
so that
\begin{equation}
N^+=2(m+n-1)
\end{equation}

When considering the negative chirality modes $\Sigma$ does not decouple.  We have $n_1=4$, $n_2=4$, $n_{z1}=3$ and $n_{z2}=3$.
Then
\begin{equation}
\begin{array}{lllllll}
\hat{q}_1^{(1)}&=q_{\chi_-}&=-q_{\psi}, & &\hat{q}_1^{(2)}&=q_{\Psi}&=1+q_{\psi}\\
\hat{q}_2^{(1)}&=q_{\tilde{\chi}_-}&=-q_{\psi},& &\hat{q}_2^{(2)}&=\tilde{q}_-^{(2)}&=1+q_{\psi}\\
\hat{q}_3^{(1)}&=\tilde{q}_{\Psi -}^{(1)}&=-q_{\psi},& &\hat{q}_3^{(2)}&=\tilde{q}_-^{(2)}&=1+q_{\psi}\\
\hat{q}_4^{(1)}&=\tilde{q}^{(1)}_{-}&=-q_{\psi},& &\hat{q}_4^{(2)}&=\tilde{q}^{(2)}_-&=1+q_{\psi}\\
\hat{q}_5^{(1)}&=\tilde{q}_{\Sigma -}^{(1)}&=1-q_{\psi},& &\hat{q}_5^{(2)}&=q_{\chi_0}&=-q_{\psi}\\
\hat{q}_6^{(1)}&=q_{\tilde{\chi}_+}&=m-q_{\psi},& &\hat{q}_6^{(2)}&=q_{\lambda}&=q_{\psi}\\
\hat{q}_7^{(1)}&=q_{\chi_+}&=n-q_{\psi},& &\hat{q}_7^{(2)}&=q_{\Sigma}&=-1+q_{\psi}\\
\end{array}
\end{equation}
so that
\begin{equation}
N^-=0
\end{equation}
In total there are $2(n+m-1)$ zero modes.  As only two of the modes vanish, this seems to reinforce the idea that there is a super-Higgs effect
occurring.

\subsection{F-term and P-term strings}
\label{zeroF}
For non-BPS strings the gravitino degrees of freedom $\Sigma$, and $\Pi$ do not decouple.  As the mass of the gravitino is non-zero the mass matrix is not block off diagonal and the index theorem
becomes more complicated than that described here.  In \cite{Brax:2006yb} it was
shown that the inclusion of gravitinos in the analysis of F-term
strings means that none of the global
SUSY zero modes survive.    Using the index theorem for generic mass
matrices gives $N^+=N^-=2$, where the extra zero
modes arise from the inclusion of the $\Pi$ gravitino.  The norm of
this field is not positive definite which suggests that these may be
gauge degrees of freedom.  If we remove the modes coming from $\Pi$ we
get $N^+=N^-=0$. Exactly the same argument applies in the semi-local case.

Apart from the D-term case all P-term models have non-vanishing
gravitino mass and the same argument that applies for F-term strings
means that there are no zero modes for a generic P-term model.  It is
the BPS nature of the D-term case which means that the zero modes
survive the coupling to supergravity.

\section{Inflation}
\label{inflation}
The P-term potential has the right vacuum structure to give hybrid
inflation.  In supersymmetry the potential has a supersymmetry
breaking vacuum when $|\phi_+|=|\phi_-|=0$ where $V=g^2\xi^2/8$.  This
is a minimum if $|\phi_0|^2>\xi/2$.  Once the critical value has been
reached the fields waterfall down into the true minimum where cosmic
strings can form. 
 We first consider a generic P-term
potential and then the special case when the FI terms appear only in
the D-term.  These two cases must be considered separately as the
scalar fields have different charges in each case.  We shall
only consider a model containing two charged chiral multiplets, as moving
to a semi-local model makes very little difference for inflation because during inflation the vevs of the charged fields are zero \cite{Urrestilla:2004eh}.

\subsection{P-term inflation}
\label{Pinflation}
If $\sin\theta\neq 0$ one of $\xi_1$,
$\xi_2$ is non-zero  and the combination $\phi_+\phi_-$ in the
superpotential (\ref{Wsugra}) must be uncharged.  We set
  $\rho_+=\rho_-=0$ in (\ref{charges}) so that
\begin{equation}
q_0=\frac{-\xi \cos\theta}{M_P^2}
\end{equation}
$\phi_0$ is the inflaton and $\phi_+$, $\phi_-$ are the waterfall
fields.  We assume that the
fields $\phi_+$ and $\phi_-$ are always much less than the Planck
mass, so neglecting terms of order $|\phi_{\pm}|^2/M_P^2$ and
higher the supergravity scalar potential becomes
\begin{eqnarray}
V&=&\frac{g^2}{2}e^{|\phi_0|^2/M_P^2}\Bigl\{|\phi_+|^2|\phi_-|^2+|\phi_0|^2|\phi_-|^2+|\phi_0|^2|\phi_+|^2\Bigr.\nonumber\\
&
&\;\;\;\;-\xi\sin\theta(e^{i\psi}\phi_+\phi_-+e^{-i\psi}\bar{\phi}_+\bar{\phi}_-)\nonumber\\
& &\;\;\;\;\left.+\xi^2\sin^2\theta\left(1-\frac{|\phi_0|^2}{M_P^2}+\frac{|\phi_0|^4}{M_P^4}\right)\right\}\nonumber\\
& &+\frac{g^2}{2}(|\phi_+|^2-|\phi_-|^2-\xi\cos\theta)\nonumber\\
& &+\frac{g^2}{2}\xi^2\cos^2\theta\frac{|\phi_0|^2}{M_P^2}\left(\frac{|\phi_0|^2}{M_P^2}+2\right)
\end{eqnarray}
The full potential is given in appendix \ref{sugrapot}.
The direction $|\phi_+|=|\phi_-|=0$ extremizes the potential in the
$\phi_+$, $\phi_-$ directions and is a minimum if
\begin{equation}
e^{2|\phi_0|^2/M_P^2}(|\phi_0|^4-4\xi^2\sin^2\theta)>4\xi^2\cos^2\theta
\end{equation}
Inflation occurs as the fields roll along this valley.
The inflationary potential is
\begin{equation}
V(|\phi_0|)=\frac{g^2\xi^2}{2}\left(1+2\frac{|\phi_0|^2}{M_P^2}\cos^2\theta+\frac{|\phi_0|^4}{2M_P^4}(1+\cos^2\theta)\right)
\end{equation}
The slow roll parameter $\eta$ is
\begin{eqnarray}
\eta&=&M_P^2\frac{V^{\prime\prime}(|\phi_0|)}{V(|\phi_0|)}\\
&=&
\frac{4\cos^2\theta+6(1+\cos^2\theta)\frac{|\phi_0|^2}{M_P^2}}{1+2\frac{|\phi_0|^2}{M_P^2}\cos^2\theta+\frac{1}{2}(1+\cos^2\theta)\frac{|\phi_0|^4}{M_P^4}}
\end{eqnarray}
There is a period of slow roll  when $|\eta| \ll 1$,  as for all
models of hybrid inflation $\epsilon
\ll \eta$.  To get slow roll requires $\cos^2\theta<1/4$, assuming that the inflaton is always less than the
Planck scale.  The
potential is bounded away from the D-term case.
Notice that $\eta$ is always strictly positive so the spectral index
$n\approx 1+ 2\eta$ is always greater than one in this model, which
disagrees with observations \cite{Spergel:2006hy}.

The one-loop corrections to the potential are
\begin{equation}
\label{oneloop}
\Delta V=\frac{\xi^2g^4}{16\pi^2}\ln\left(\frac{|\phi_0|^2}{\Lambda}\right)
\end{equation}
where $\Lambda$ is a symmetry breaking scale.  The slow roll equations of
motion for the resulting effective potential are
\begin{equation}
H^2\approx\frac{g^2\xi^2}{6M_P^2}
\end{equation}
\begin{equation}
3H\frac{|\dot{\phi}_0|}{M_P} =\frac{-g^2\xi^2}{2}\left(\frac{g^2M_P}{4\pi^2|\phi_0|} +4\frac{|\phi_0|}{M_P}
\cos^2\theta +2\frac{|\phi_0|^3}{M_P^3}(1+\cos^2\theta)\right)
\end{equation}
and we require $N=60$  efolds of inflation to agree with
observations.  The solutions to these equations fall into two classes depending on the relative
values of $g$ and $\theta$.

 If $(1+\cos^2\theta)g^2<8\pi^2\cos^4\theta$ then
\begin{equation}
\label{x_N1}
\frac{|\phi_0|_N^2}{M_P^2}=\frac{A(B+e^{8A(1+\cos^2\theta)N})}{B-e^{8A(1+\cos^2\theta)N}}-\frac{\cos^2\theta}{1+\cos^2\theta}
\end{equation}
where
\begin{equation}
A^2 =
\frac{8\pi^2\cos^4\theta-g^2(1+\cos^2\theta)}{8\pi^2(1+\cos^2\theta)^2}
\end{equation}
and assuming $|\phi_0|_{\tiny \mbox{end}} \ll |\phi_0|_N$ 
\begin{equation}
B \approx
\frac{\cos^2\theta+(1+\cos^2\theta)A}{\cos^2\theta-(1+\cos^2\theta)A}
\end{equation}
To ensure $|\phi_0|_N^2$ is non-negative in (\ref{x_N1}) requires
\begin{equation}
8\cos^2\theta\left(1-\frac{(1+\cos^2\theta)g^2}{16\pi^2\cos^4\theta}\right)N<1
\end{equation}
which restricts $\theta$ and $g$ in the following way
\begin{equation}
\cos^2\theta<4 \times 10^{-3}
\end{equation}
\begin{equation}
g^2<4 \times 10^{-4}
\end{equation}

Alternatively if $(1+\cos^2\theta)g^2>8\pi^2\cos^4\theta$ then
\begin{equation}
\frac{|\phi_0|_N^2}{M_P^2}=\frac{-\cos^2\theta}{1+\cos^2\theta}+A\tan\left(4N(1+\cos^2\theta)A+\arctan\left(\frac{\cos^2\theta}{A(1+\cos^2\theta)}\right)\right)
\end{equation}
assuming  $|\phi_0|_{\tiny \mbox{end}}\ll |\phi_0|_N$.  
Insisting that this is single valued gives
\begin{equation}
g^2\leq\frac{\frac{\pi^4}{8N^2}+8\pi^2\cos^2\theta}{1+\cos^2\theta}
\end{equation}
so $0\leq g^2\leq 15.8$.

The COBE normalisation for the density perturbations at horizon
crossing \cite{Lyth:1998xn} is
\begin{equation}
\frac{1}{5\sqrt{3}\pi}\frac{V^{3/2}(|\phi_0|_N)}{V^{\prime}(|\phi_0|_N)} \sim 1.9
\times 10^{-5}
\end{equation}
If $g^2(1+\cos^2\theta)<8\pi^2\cos^4\theta$ this requires $\xi \gtrsim
4 \times 10^{-5}$, and if $g^2 (1+\cos^2\theta)>8\pi^2\cos^4\theta$ then $\xi \gtrsim 3.4 \times 10^{-6}$.
 $\xi$ is always too large to agree with the bounds  on the string
tension discussed in section \ref{NO}.  Hence NO strings cannot form at
the end of inflation, but semi-local strings are allowed as they
are not stable on cosmological timescales.
\subsection{The D-term Case}
\label{Dinflation}
If $\sin\theta=0$ then there are no FI terms in the superpotential and
the combination $\phi_+\phi_-$ no longer has to be uncharged;
$q_{\pm}\neq Q_{\pm}$. The resulting inflationary potential depends on the
parameter $q_0=-\rho_0\xi/M_P^2$.

The tree level inflationary potential is
\begin{equation}
V(\phi_0)=\frac{g^2\xi^2}{2}\left(\frac{\rho_0|\phi_0|^2}{M_P^2}+1
\right)^2
\end{equation}
 and there exists a period of slow roll with $|\eta|\ll 1$, if $|\rho_0| <1/4$.
The loop corrections are as in (\ref{oneloop}), so that the slow roll equations are
\begin{equation}
H^2 \approx \frac{g^2 \xi^2}{6M_P^2}
\end{equation}
\begin{equation}
3H\frac{|\dot{\phi}_0|}{M_P}=\frac{-g^2 \xi^2}{2}\left(4\rho_0^2\frac{|\phi_0|^3}{M_P^3}+4\rho_0 \frac{|\phi_0|}{M_P}
+\frac{g^2M_P}{4\pi^2 |\phi_0|}\right)
\end{equation}
If we
make the reasonable assumption $g^2<4\pi^2$ these admit the same form
of solution as the first case considered above.  For $|\phi_0|^2_N$ to
be non-negative requires $\rho_0<0$.  $-1/4<\rho_0<0$ means that the
slow roll parameter 
\begin{equation}
\eta =
\frac{4\rho_0\left(1+\frac{3\rho|\phi_0|^2}{M_P^2}\right)}{\left(1+\frac{\rho_0|\phi_0|^2}{M_P^2}\right)^2}
\end{equation}
is always negative so the spectral index is less than one in agreement
with observations \cite{Spergel:2006hy}.  For the density perturbations to be in agreement with observations requires
\begin{equation}
\xi \gtrsim 4.5 \times 10^{-4}
\end{equation}
Which is again too high to agree with bounds on the string tension for
NO strings. However, semi-local strings could form in the two doublet model,
as in \cite{Urrestilla:2004eh}.

\subsection{String Formation at the End of Inflation}
Inflation ends either when the slow roll conditions are violated, or
when the field is no longer rolling in a valley.  When the fields
leave the valley they waterfall down into the supersymmetric
vacuum, where NO or semi-local strings may form depending on the
choice of model.  

It is well known that if NO strings form at the end of hybrid
inflation the tension of the strings is typically too high to agree
with observations.  However in \cite{Rocher:2004uv,Sakellariadou:2004rf,Rocher:2004et} it was shown that for F-term and
D-term supersymmetry  when the
superpotential coupling $\beta$  and coupling constant $g$ were detuned,
additional radiative corrections were considered and the fields were
allowed to roll near the Planck scale there was a region of
parameter space which allowed both sufficient inflation and NO cosmic
strings.  It is probable that a similar analysis when applied to P-term
inflation and cosmic strings would find that there was a region of
parameter space where they were both permitted.  However, this detuning 
would destroy the underlying supersymmetry. 
Moving to a semi-local model so that topologically unstable strings form at
the end of inflation means that we can avoid the string tension
constraints without detuning the couplings, thus keeping the underlying
supersymmetry of the theory. In \cite{Jeannerot:2005mc} the conflict
between inflation and the resulting cosmic strings was avoided by
detuning the couplings and by considering the warm inflation and
curvaton scenarios, but again this would mean breaking the underlying
symmetries of the model.

It is possible that the FI terms arise as vevs of fields which are
fixed in a compactification scheme \cite{Achucarro:2006zf}, in which case we do not need to
alter the charges of the scalar fields.  However in this case the bounds on $\xi$ are
still too large to allow NO strings to be formed at the end of inflation.

\section{Conclusions}
\label{conc}
P-term potentials have the right vacuum structure to give hybrid
inflation and cosmic strings, however NO cosmic strings have a
tension which is too high to agree with observations of inflation.  Moving to a
semi-local model means that the strings that form are unstable and do
not conflict with cosmological observations.

We have given the general solutions to the Bogomol'nyi equations for both NO and semi-local
strings in P-term potentials and examined the behaviour of the fields
at large and small distances from the string.  All P-term strings have
the same energy, which means that current estimates of the string
tension put the same bound on the FI term in all P-term models.
However this is not a problem for semi-local strings because of their instability.

D-term strings are BPS states whereas all other forms of P-term
strings break all of the supersymmetries.  This means that in an analysis
of the string zero modes the D-term case must be treated separately
from the generic P-term case.  In supergravity D-term NO
strings have $2(n-1)$ zero modes and D-term semi-local strings have
$2(n+m-1)$ zero modes.  When compared to the supersymmetry results of
$2n$ and $2(n+m)$ respectively this seems to indicate that a super
Higgs effect is occurring.  For all other types of cosmic string
forming in P-term potentials no zero modes survive the move from supersymmetry to supergravity.  

We expect these cosmic strings to form at the end of a period of
hybrid inflation.  The charges of the scalar fields differ in
supergravity from the supersymmetry values, which required a
re-analysis of inflation in P-term potentials.  It was found that
the value of $\xi$ required to give density perturbations of the right
order to agree with observations is too high to agree with the string
tension bound for NO strings.  This makes the semi-local model
preferable.  In addition we note that only the D-term model of
inflation gives a spectral index which is less than one in agreement
with observations.

\section*{Acknowledgements}
This work was partially supported by PPARC.

\appendix
\section{Euler Angles}
\label{angles}
We use the following parameterisations for an SO(3) rotation in terms
of Cayley-Klein parameters
\begin{equation}
R=\left(\begin{array}{ccc}
\frac{1}{2}(a^2-b^{*2}+a^{*2}-b^2) &
\frac{i}{2}(b^{*2}-a^2+a^{*2}-b^2) & -ab-a^*b^*\\
\frac{i}{2}(a^2+b^{*2}-a^{*2}-b^2) &
\frac{1}{2}(b^{*2}+a^2+a^{*2}+b^2) & -i(ab-a^*b^*)\\
ba^*+ab^* & i(ba^*-ab^*) & aa^*-bb^*
\end{array}\right)
\end{equation}
which are defined in terms of Euler angles as
\begin{equation}
a=e^{i(\phi+\psi)/2}\cos\frac{\theta}{2}
\end{equation}
\begin{equation}
b=ie^{i(\phi-\psi)/2}\sin\frac{\theta}{2}
\end{equation}
The SU(2) rotation associated with this SO(3) rotation is 
\begin{equation}
U=\left(\begin{array}{cc}
a & b\\
-b^* & a^*
\end{array}\right)
\end{equation}

\section{P-term Supergravity Potential}
\label{sugrapot}
the full supergravity potential can be calculated form 
\begin{equation}
V=e^K\left(\left|\frac{\partial W}{\partial
  \phi_i}+\frac{\phi_i^*W}{M_P^2}\right|^2-\frac{3|W|^2}{M_P^2}\right)+D^2
\end{equation}
With
superpotential (\ref{Wsugra}) and D-term (\ref{Dsugra}) the supergravity potential for bosonic scalars
is 
\begin{eqnarray}
V=&
&\frac{g^2}{2}e^K\left\{|\phi_+\phi_-|^2\left(1+\frac{|\phi_0|^4}{M_P^4}\right)+|\phi_0\phi_-|^2\left(1+\frac{|\phi_+|^4}{M_P^4}\right)\right.\nonumber\\
&
&\;\;\;\;+|\phi_0\phi_+|^2\left(1+\frac{|\phi_-|^4}{M_P^4}\right)+\frac{3|\phi_0\phi_+\phi_-|^2}{M_P^2}\nonumber\\
& &\;\;\;\;-\sin\theta\xi(e^{i\psi}\phi_+\phi_-+e^{-i\psi}\bar{\phi}_+\bar{\phi}_-)\nonumber\\
&
&\;\;\;\;\;\;\;\;\times\left(1+\frac{|\phi_0|^2}{M_P^2}+\frac{|\phi_0|^2}{M_P^4}(|\phi_0|^2+|\phi_+|^2+|\phi_-|^2)\right)\nonumber\\
& &\;\;\;\;+(\sin\theta\xi)^2\nonumber\\
& &\;\;\;\;\;\;\;\;\left.\times\left(1-\frac{|\phi_0|^2}{M_P^2}+\frac{|\phi_0|^2}{M_P^4}(|\phi_0|^2+|\phi_+|^2+|\phi_-|^2)\right)\right\}\nonumber\\
&+&\frac{g^2}{2}\left(-\frac{\xi\cos\theta}{M_P^2}|\phi_0|^2+|\phi_+|^2-|\phi_-|^2-\cos\theta\xi\right)^2
\end{eqnarray}

\bibliography{reportbib}

\providecommand{\href}[2]{#2}\begingroup\raggedright\begin{thebibliography}{10}

\bibitem{Kachru:2003aw}
S.~Kachru, R.~Kallosh, A.~Linde, and S.~P. Trivedi, {\it De {S}itter vacua in
  string theory},  {\em Phys. Rev.} {\bf D68} (2003) 046005,
  [\href{http://xxx.lanl.gov/abs/hep-th/0301240}{{\tt hep-th/0301240}}].

\bibitem{Kachru:2003sx}
S.~Kachru, R.~Kallosh, A.~Linde, J.~Maldacena, L.~McAllister, and S.~P.
  Trivedi, {\it Towards inflation in string theory},  {\em JCAP} {\bf 0310}
  (2003) 013, [\href{http://xxx.lanl.gov/abs/hep-th/0308055}{{\tt
  hep-th/0308055}}].

\bibitem{Majumdar:2002hy}
M.~Majumdar and A.-C. Davis, {\it Cosmological creation of {D}-branes and
  anti-{D}-branes},  {\em JHEP} {\bf 03} (2002) 056,
  [\href{http://xxx.lanl.gov/abs/hep-th/0202148}{{\tt hep-th/0202148}}].

\bibitem{Sarangi:2002yt}
S.~Sarangi and S.~H.~H. Tye, {\it Cosmic string production towards the end of
  brane inflation},  {\em Phys. Lett.} {\bf B536} (2002) 185--192,
  [\href{http://xxx.lanl.gov/abs/hep-th/0204074}{{\tt hep-th/0204074}}].

\bibitem{Polchinski:2004ia}
J.~Polchinski, {\it Introduction to cosmic {F}- and {D}-strings},
  \href{http://xxx.lanl.gov/abs/hep-th/0412244}{{\tt hep-th/0412244}}.

\bibitem{Davis:2005dd}
A.~C. Davis and T.~W.~B. Kibble, {\it Fundamental cosmic strings},  {\em
  Contemp. Phys.} {\bf 46} (2005) 313--322,
  [\href{http://xxx.lanl.gov/abs/hep-th/0505050}{{\tt hep-th/0505050}}].

\bibitem{Dasgupta:2004dw}
K.~Dasgupta, J.~P. Hsu, R.~Kallosh, A.~Linde, and M.~Zagermann, {\it {D}3/{D}7
  brane inflation and semilocal strings},  {\em JHEP} {\bf 08} (2004) 030,
  [\href{http://xxx.lanl.gov/abs/hep-th/0405247}{{\tt hep-th/0405247}}].

\bibitem{Dasgupta:2002ew}
K.~Dasgupta, C.~Herdeiro, S.~Hirano, and R.~Kallosh, {\it D3/{D}7 inflationary
  model and {M}-theory},  {\em Phys. Rev.} {\bf D65} (2002) 126002,
  [\href{http://xxx.lanl.gov/abs/hep-th/0203019}{{\tt hep-th/0203019}}].

\bibitem{Kallosh:2001tm}
R.~Kallosh, {\it {N} = 2 supersymmetry and de {S}itter space},
  \href{http://xxx.lanl.gov/abs/hep-th/0109168}{{\tt hep-th/0109168}}.

\bibitem{Kallosh:2003ux}
R.~Kallosh and A.~Linde, {\it P-term, {D}-term and {F}-term inflation},  {\em
  JCAP} {\bf 0310} (2003) 008,
  [\href{http://xxx.lanl.gov/abs/hep-th/0306058}{{\tt hep-th/0306058}}].

\bibitem{Burrage:2007yu}
C.~Burrage and A.~C. Davis, {\it P-term potentials from 4-{D} supergravity},
  {\em JHEP} {\bf 06} (2007) 086,
  [\href{http://xxx.lanl.gov/abs/arXiv:0705.1657 [hep-th]}{{\tt arXiv:0705.1657
  [hep-th]}}].

\bibitem{Freedman:1977}
D.~Freedman, {\it Supergravity with {A}xial {G}auge {I}nvariance},  {\em Phys.
  Rev.} {\bf D15} (1977) 1173.

\bibitem{Stelle:1978wj}
K.~S. Stelle and P.~C. West, {\it Relation between vector and scalar multiplets
  and gauge invariance in supergravity},  {\em Nucl. Phys.} {\bf B145} (1978)
  175.

\bibitem{Barbieri:1982ac}
R.~Barbieri, S.~Ferrara, D.~V. Nanopoulos, and K.~S. Stelle, {\it Supergravity,
  {R} invariance and spontaneous supersymmetry breaking},  {\em Phys. Lett.}
  {\bf B113} (1982) 219.

\bibitem{Binetruy:2004hh}
P.~Binetruy, G.~Dvali, R.~Kallosh, and A.~Van~Proeyen, {\it Fayet-{I}liopoulos
  terms in supergravity and cosmology},  {\em Class. Quant. Grav.} {\bf 21}
  (2004) 3137--3170, [\href{http://xxx.lanl.gov/abs/hep-th/0402046}{{\tt
  hep-th/0402046}}].

\bibitem{Kibble:1976sj}
T.~W.~B. Kibble, {\it Topology of cosmic domains and strings},  {\em J. Phys.}
  {\bf A9} (1976) 1387--1398.

\bibitem{Brax:2006yb}
P.~Brax, C.~van~de Bruck, A.~C. Davis, and S.~C. Davis, {\it Fermionic zero
  modes of supergravity cosmic strings},  {\em JHEP} {\bf 06} (2006) 030,
  [\href{http://xxx.lanl.gov/abs/hep-th/0604198}{{\tt hep-th/0604198}}].

\bibitem{Achucarro:2004ry}
A.~Achucarro and J.~Urrestilla, {\it F-term strings in the {B}ogomolnyi limit
  are also {B}{P}{S} states},  {\em JHEP} {\bf 08} (2004) 050,
  [\href{http://xxx.lanl.gov/abs/hep-th/0407193}{{\tt hep-th/0407193}}].

\bibitem{Davis:1997bs}
S.~C. Davis, A.~C. Davis, and M.~Trodden, {\it N = 1 supersymmetric cosmic
  strings},  {\em Phys. Lett.} {\bf B405} (1997) 257--264,
  [\href{http://xxx.lanl.gov/abs/hep-ph/9702360}{{\tt hep-ph/9702360}}].

\bibitem{Linde}
A.~Linde, {\it Axions in inflationary cosmology},  {\em Phys. Lett. B} {\bf
  259} (1991) 38.

\bibitem{Achucarro:2005vz}
A.~Achucarro, A.~Celi, M.~Esole, J.~Van~den Bergh, and A.~Van~Proeyen, {\it
  D-term cosmic strings from {N} = 2 supergravity},  {\em JHEP} {\bf 01} (2006)
  102, [\href{http://xxx.lanl.gov/abs/hep-th/0511001}{{\tt hep-th/0511001}}].

\bibitem{VilShel}
A.~Vilenkin and E.~P.~S. Shellard, {\em Cosmic Strings and Other Topological
  Defects}.
\newblock Cambridge University Press, 1994.

\bibitem{Comtet:1987wi}
A.~Comtet and G.~W. Gibbons, {\it Bogomolny bounds for cosmic strings},  {\em
  Nucl. Phys.} {\bf B299} (1988) 719.

\bibitem{Copeland:2003bj}
E.~J. Copeland, R.~C. Myers, and J.~Polchinski, {\it Cosmic {F}- and
  {D}-strings},  {\em JHEP} {\bf 06} (2004) 013,
  [\href{http://xxx.lanl.gov/abs/hep-th/0312067}{{\tt hep-th/0312067}}].

\bibitem{Hindmarsh:1992yy}
M.~Hindmarsh, {\it Semilocal topological defects},  {\em Nucl. Phys.} {\bf
  B392} (1993) 461--492, [\href{http://xxx.lanl.gov/abs/hep-ph/9206229}{{\tt
  hep-ph/9206229}}].

\bibitem{Penin:1996si}
A.~A. Penin, V.~A. Rubakov, P.~G. Tinyakov, and S.~V. Troitsky, {\it What
  becomes of vortices in theories with flat directions},  {\em Phys. Lett.}
  {\bf B389} (1996) 13--17, [\href{http://xxx.lanl.gov/abs/hep-ph/9609257}{{\tt
  hep-ph/9609257}}].

\bibitem{Achucarro:1999it}
A.~Achucarro and T.~Vachaspati, {\it Semilocal and electroweak strings},  {\em
  Phys. Rept.} {\bf 327} (2000) 347--426,
  [\href{http://xxx.lanl.gov/abs/hep-ph/9904229}{{\tt hep-ph/9904229}}].

\bibitem{Urrestilla:2004eh}
J.~Urrestilla, A.~Achucarro, and A.~C. Davis, {\it D-term inflation without
  cosmic strings},  {\em Phys. Rev. Lett.} {\bf 92} (2004) 251302,
  [\href{http://xxx.lanl.gov/abs/hep-th/0402032}{{\tt hep-th/0402032}}].

\bibitem{Jeannerot:2004bt}
R.~Jeannerot and M.~Postma, {\it Chiral cosmic strings in supergravity},  {\em
  JHEP} {\bf 12} (2004) 043,
  [\href{http://xxx.lanl.gov/abs/hep-ph/0411260}{{\tt hep-ph/0411260}}].

\bibitem{Spergel:2006hy}
D.~N. Spergel {\em et~al.}, {\it Wilkinson {M}icrowave {A}nisotropy {P}robe
  ({WMAP}) three year results: {I}mplications for cosmology},
  \href{http://xxx.lanl.gov/abs/astro-ph/0603449}{{\tt astro-ph/0603449}}.

\bibitem{Lyth:1998xn}
D.~H. Lyth and A.~Riotto, {\it Particle physics models of inflation and the
  cosmological density perturbation},  {\em Phys. Rept.} {\bf 314} (1999)
  1--146, [\href{http://xxx.lanl.gov/abs/hep-ph/9807278}{{\tt
  hep-ph/9807278}}].

\bibitem{Rocher:2004uv}
J.~Rocher and M.~Sakellariadou, {\it Consistency of cosmic strings with cosmic
  microwave background measurements},
  \href{http://xxx.lanl.gov/abs/hep-ph/0405133}{{\tt hep-ph/0405133}}.

\bibitem{Sakellariadou:2004rf}
M.~Sakellariadou and J.~Rocher, {\it Constraining {SUSY} {GUT}s with
  cosmology},  \href{http://xxx.lanl.gov/abs/hep-ph/0406164}{{\tt
  hep-ph/0406164}}.

\bibitem{Rocher:2004et}
J.~Rocher and M.~Sakellariadou, {\it Supersymmetric grand unified theories and
  cosmology},  {\em JCAP} {\bf 0503} (2005) 004,
  [\href{http://xxx.lanl.gov/abs/hep-ph/0406120}{{\tt hep-ph/0406120}}].

\bibitem{Jeannerot:2005mc}
R.~Jeannerot and M.~Postma, {\it Confronting hybrid inflation in supergravity
  with cmb data},  {\em JHEP} {\bf 05} (2005) 071,
  [\href{http://xxx.lanl.gov/abs/hep-ph/0503146}{{\tt hep-ph/0503146}}].

\bibitem{Achucarro:2006zf}
A.~Achucarro, B.~de~Carlos, J.~A. Casas, and L.~Doplicher, {\it de {S}itter
  vacua from uplifting {D}-terms in effective supergravities from realistic
  strings},  {\em JHEP} {\bf 06} (2006) 014,
  [\href{http://xxx.lanl.gov/abs/hep-th/0601190}{{\tt hep-th/0601190}}].

\end{thebibliography}\endgroup
\end{document}